\documentclass[10pt,emptycopyrightspace]{ewsn-proc}

\usepackage{graphicx}
\usepackage{balance}
\usepackage{comment}
\usepackage{pbox}
\usepackage{amsmath}
\usepackage{algorithm, algorithmic}
\usepackage{subfig}
\usepackage[justification=centering]{caption}
\usepackage{makecell}

\usepackage{amsfonts}
\usepackage{xcolor}
\usepackage{url}

\title{Hermes: Decentralized Dynamic Spectrum Access System for Massive Devices Deployment in 5G}

\numberofauthors{5}
\author{
\alignauthor {Zhihui Gao} \\
    \affaddr{Duke University}\\
    \email{zhihui.gao@duke.edu}
    \and
\alignauthor {Ang Li} \\
    \affaddr{Duke University}\\
    \email{ang.li630@duke.edu}
    \and
\alignauthor {Yunfan Gao} \\
    \affaddr{ETH Zurich}\\
    \email{yungao@student.ethz.ch}
  \and
\alignauthor {Yu Wang} \\
    \affaddr{Tsinghua University}\\
    \email{yu-wang@tsinghua.edu.cn}
   \and
\alignauthor {Yiran Chen} \\
    \affaddr{Duke University}\\
    \email{yiran.chen@duke.edu}
}

\begin{document}

\maketitle

\begin{abstract}
With the incoming 5G network, the ubiquitous Internet of Things (IoT) devices can benefit our daily life, such as smart cameras, drones, etc. With the introduction of the millimeter-wave band and the thriving number of IoT devices, it is critical to design new dynamic spectrum access (DSA) system to coordinate the spectrum allocation across massive devices in 5G.
In this paper, we present Hermes, the first decentralized DSA system for massive devices deployment. Specifically, we propose an efficient multi-agent reinforcement learning algorithm and introduce a novel shuffle mechanism, addressing the drawbacks of collision and fairness in existing decentralized systems.
We implement Hermes in 5G network via simulations. Extensive evaluations show that Hermes significantly reduces collisions and improves fairness compared to the state-of-the-art decentralized methods.
Furthermore, Hermes is able to adapt the environmental changes within 0.5 seconds, showing its deployment practicability in dynamic environment of 5G.

\end{abstract}

\category{Computer Systems Organization}{COMPUTER-COMMUNICATION NETWORKS}{Network Architecture and Design}
\terms{Design, Standardization}

\section{Introduction}
\label{sec:introduction}

Recent development of the fifth-generation (5G) network as well as the explosive growth of the Internet of Things (IoT) devices draw attention to spectrum management (SM), especially dynamic spectrum access.
The hybrid spectrum landscape in 5G, i.e., microwave and millimeter-wave (mmWave) bands, provides more available spectrum resources than before~\cite{niknam2020federated}.
In addition, the thriving numbers of the ubiquitous IoT devices emerge in our daily life, such as smartwatches, augmented/virtual reality headsets, and self-driving cars. With the increase of available spectrum and IoT devices, an urgent demand is required on an efficient and fair spectrum allocation management.

The traditional SM refers to centralized methods that are deployed on a central processor, such as a base station. The central processor collects all the sensory data from multiple user equipments (UEs) and schedules how to allocate the limited channels to UEs. The channels allocated by the central processor are referred to as the licensed channels.
In the cognitive networks, besides the licensed channels, UEs are also able to opportunistically access the temporarily unused channels or unlicensed channels, which is termed as the dynamic spectrum access (DSA)~\cite{akyildiz2006next,akyildiz2008survey}.
DSA often adopts decentralized methods that are deployed on UEs, such that UEs can access the shared spectrum coordinately. 
Compared to the centralized SM, {\color{black}UEs do not have to wait for the decision by the base station but decides their action by their own. Hence, the response delay is greatly decreased in DSA.} However, it is very challenging for multiple UEs to coordinate a schedule plan without collisions.

\begin{figure}[t]
    \centering
    \includegraphics[width=0.8\columnwidth]{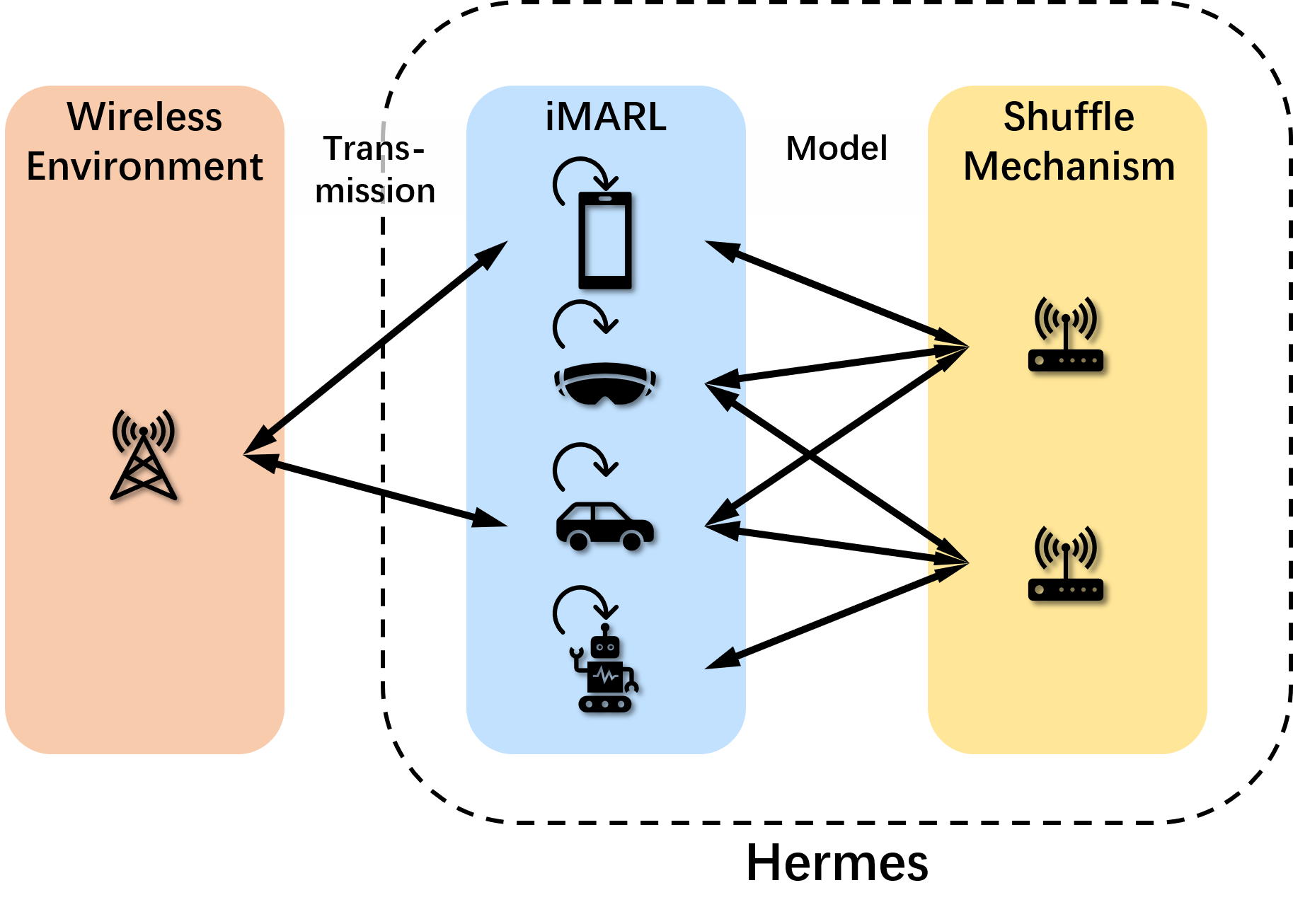}
    \caption{A high-level overview of Hermes.}
    \label{fig: overview}
\end{figure}

One of the most widely used SM methods in 4G/5G cellular network is the proportional fairness (PF)~\cite{tiwari2014long}. PF has a better balance between the total throughput and fairness compared to other methods such as round robin and best CQI {\color{black}(channel quality indication)}. As a centralized method, PF is also bottlenecked by the serious delay when the number of UEs increases. In addition, all the sensory data of UEs needs to be uploaded to a central processor, which raises privacy concerns~\cite{bahl2000radar,zhu2018tu}.
As for the decentralized method in DSA, UEs make their decisions independently to maximize their own benefits. Therefore, game theory is introduced to analyze the scheduling strategy and benefits of UEs. For example, the game theory based algorithm ALLURE-U~\cite{vamvakas2019dynamic} provides an optimal transmission plan in a realistic setting. The plan decides how much power to allocate over licensed and unlicensed channels. The prospect theory is exploited to estimate the Nash Equilibrium. However, the output of ALLURE-U is only heuristical. It does not generate a detailed scheduling plan for each time slot. Therefore, it cannot be directly applied.
Besides game theory, multi-agent reinforcement learning (MARL) and deep Q network (DQN)~\cite{mnih2015human} are applied in DSA. The deep Q-learning spectrum access (DQSA)~\cite{naparstek2018deep} deploys DQN on UEs and enables UEs to choose their own channels without a central processor or sharing their sensory data. DQSA works well when there are enough channels for each UE. However, as the number of UEs increases, DQSA fails in two aspects. First, DQSA shows poor fairness because UEs do not share the channel alternatively but occupy certain channels. It is even worse that UEs are so aggressive that UEs would rather allow collisions to occur than tolerating other UEs' occupying channels alone without collisions.

\begin{table}[ht]
    \centering
    \caption{Comparison between Hermes and existing SM and DSA techniques.}
    \begin{tabular}{lccc}
        \hline
        \textbf{Methods} & \textbf{Mode} &
        \textbf{Fairness}& \makecell[c]{\textbf{UE}\\ \textbf{Scalability}}\\
        \hline
        PF & Centralized & Yes & No\\
        ALLURE-U & - & No & Yes\\
        DQSA & Decentralized & No & No\\
        \textbf{Hermes} & \textbf{Decentralized} & \textbf{Yes} & \textbf{Yes}\\
        \hline
    \end{tabular}
    \label{tab: Comparison}
\end{table}

In this paper, we propose \textit{Hermes} -- a decentralized DSA system in 5G network that can effectively address the aforementioned drawbacks of existing MARL such as DQSA and achieve comparable performance with centralized PF. As Fig. \ref{fig: overview} shows, Hermes consists of two modules: improved multi-agent reinforcement learning (iMARL) that is deployed on each UE, and a shuffle mechanism that is deployed on local transceivers as shufflers.
These two modules can overcome the challenges of collisions and fairness respectively. In iMARL, we define a novel reward function for the Q-learning such that the Nash Equilibrium changes and collisions are significantly reduced. In addition, two modifications to DQN's workflow are made to adapt the shuffle mechanism. The novel shuffle mechanism, deployed on single or multiple shufflers, is proposed to improve fairness. Each shuffler does not need to collect all the models but a portion from participating UEs. With the received models, the shuffler evaluates the UEs' preference to each model and distributes models back accordingly. As the models are shuffled, the UEs' selections for channels are shuffled. Hence, UEs can fairly share all channels. Note that during a shuffle only the parameters of iMARL model are uploaded, and hence the privacy of sensory data is protected.

Table \ref{tab: Comparison} provides a comparison between Hermes and existing SM and DSA techniques. Hermes differs from PF as it is a decentralized solution that offers high UE scalability. Compared to ALLURE-U, Hermes provides a practical scheduling plan for each UE, taking fairness into account. In addition, Hermes overcomes the drawbacks of fairness and UE scalability in DQSA.

We summarize four major contributions of Hermes:
\begin{itemize}
    \item To the best of our knowledge, Hermes is the first decentralized DSA system that enables UEs to efficiently and fairly share channels in massive UE deployment.
    \item An improved multi-agent reinforcement learning is proposed that significantly reduce the channel collisions. Besides, iMARL contains a compact DQN structure specifically designed for Hermes, such that the communication cost of sharing the model with other UEs is significantly saved.
    \item We proposed a novel shuffle mechanism that shares the models from multiple UEs to achieve better fairness with the UEs' privacy protected.
    \item We implement a prototype system of Hermes, and conduct extensive experiments via simulating various 5G settings. Experiment results demonstrate that Hermes significantly reduces the collisions and improves fairness compared to the state-of-the-art decentralized techniques.
\end{itemize}
\section{Background and Motivation}\label{sec:background}
We begin with introducing the existing centralized SM and decentralized DSA techniques. We then show that the state-of-the-art MARL based approach incurs both serious collisions and unfairness when allocating channels to UEs, which motivates the design of Hermes.
\subsection{Spectrum Management and Centralized Method in 5G}
Spectrum management (SM)~\cite{akyildiz2008survey} in cellular network refers to the process that a central base station allocates and schedules the limited channels on the spectrum to multiple UEs over time. The goal of SM is to maximize total throughput while guaranteeing fairness over UEs.
To achieve this goal in terms of 5G, the scheduling is performed in three steps in a period.
First, all the UEs are required to upload their sensory data to a 5G new radio gNodeB (gNb, i.e., the base station in 5G). The typical sensory data includes channel quality indication (CQI) that describes the throughput capacity of each channel.
Second, gNb runs a scheduling strategy, such as PF, to determine the scheduling plan for the current periodicity based on the received sensory data from all the UEs.
Finally, the scheduling plan is offloaded to UEs and is executed. UEs can transmit data in the allocated channels until the next periodicity.

Such a centralized method guarantees that UEs work cooperatively without collision, and hence yields good performance. However, it leads to some privacy issues. For example, the absolute value of CQI indicates the distance from the UE to the gNb~\cite{bahl2000radar}, such information may breach UEs' localization and mobility privacy~\cite{zhu2018tu}.

In addition, with the explosive increase of IoT devices {\color{black}(over 1 million connected devices per square kilometers\cite{device5G})} that require massive channels, it is not practical for a gNb to efficiently manage all the UEs simultaneously. The huge number of UEs is more challenging to synchronously upload sensory data on time and it takes longer time to perform the scheduling. Hence, the latency of SM dramatically increases. {\color{black}For example, Verizon reported the latency of 5G in 2019 is 30 milliseconds~\cite{latency5G}}
Furthermore, with the constrained spectrum resources, uploading sensory data and offloading schedule plans incur non-negligible resource consumption when considering massive UEs are involved.

\subsection{Dynamic Spectrum Access}
Compared to SM, DSA~\cite{zhao2007survey} is a decentralized mechanism developed for cognitive radios. Being cognitive means that radios can automatically detect available channels and choose the best one to use.
In practice, gNb connects to a portion of licensed UEs such as smartphones, namely primary users (PUs), and runs the centralized scheduling strategy for allocating channels to PUs.
The rest of UEs are considered as unlicensed secondary users (SUs), they need to develop the scheduling plan independently and opportunistically under the principle that the PUs' communication channels are not interfered. For example, an SU can occupy a temporary idle channel that is not allocated to any PUs or transmit with low power, which does not interfere PUs' communication. SUs can be any kind of IoT devices other than the PUs, such as smartwatches and glasses, augmented/virtual reality headsets, etc.

\subsection{Reinforcement Learning}
To address DSA at the SU side, reinforcement learning (RL) based methods are proposed. The typical RL-based DSA is single agent paradigm~\cite{wang2019survey, morozs2015distributed, morozs2015heuristically} that deploys RL models on a single SU. The goal of RL is to learn the transmission pattern of nearby PUs and minimize the SU's collision with them.
RL can be formulated with three components: state, action and reward. In this RL methods, the state includes the UE's own communication requirement and the sensed signal to interference and noise ratio (SINR) of each available channel, which is fully perceived by the UE. The UE decides whether to transmit or which channel to transmit as the action. It gets positive rewards if data is transmitted successfully and negative if not.

\subsection{Multi-Agent Reinforcement Learning}\label{sec:back_MARL}
In most cases, more than one SUs share the unlicensed channels that are not occupied by PUs, where the interference among SUs needs to be considered. Therefore, multi-agent reinforcement learning (MARL)~\cite{naparstek2018deep} based methods are proposed.
MARL focuses on the competition and cooperation among SUs, ignoring the existence of PUs. In particular, MARL-based method assumes a mapping from the non-consecutive unlicensed channels to consecutive virtual channels as if there are no PUs in these virtual channels and DSA is performed on the consecutive virtual channels. In this case, collision (i.e., two or more UEs choose the same channel but none of them is able to successfully transmit) and idle channel (none of the UEs choose a certain channel) are two critical challenges.

Note that each UE can only perceive its own state instead of the states of peers in MARL. This means the observation that UE obtains cannot fully represent the state. In such a non-Markovian model, UEs are supposed to decide their next action based on not only the current observation, but also the action history in recent time slots as additional observations. With more information added to the state vector, the state space is exponentially extended for tabular Q learning. Hence, deep Q network (DQN) is introduced to assist the decision.

\begin{figure}[t]
    \centering
    \includegraphics[width=0.8\columnwidth]{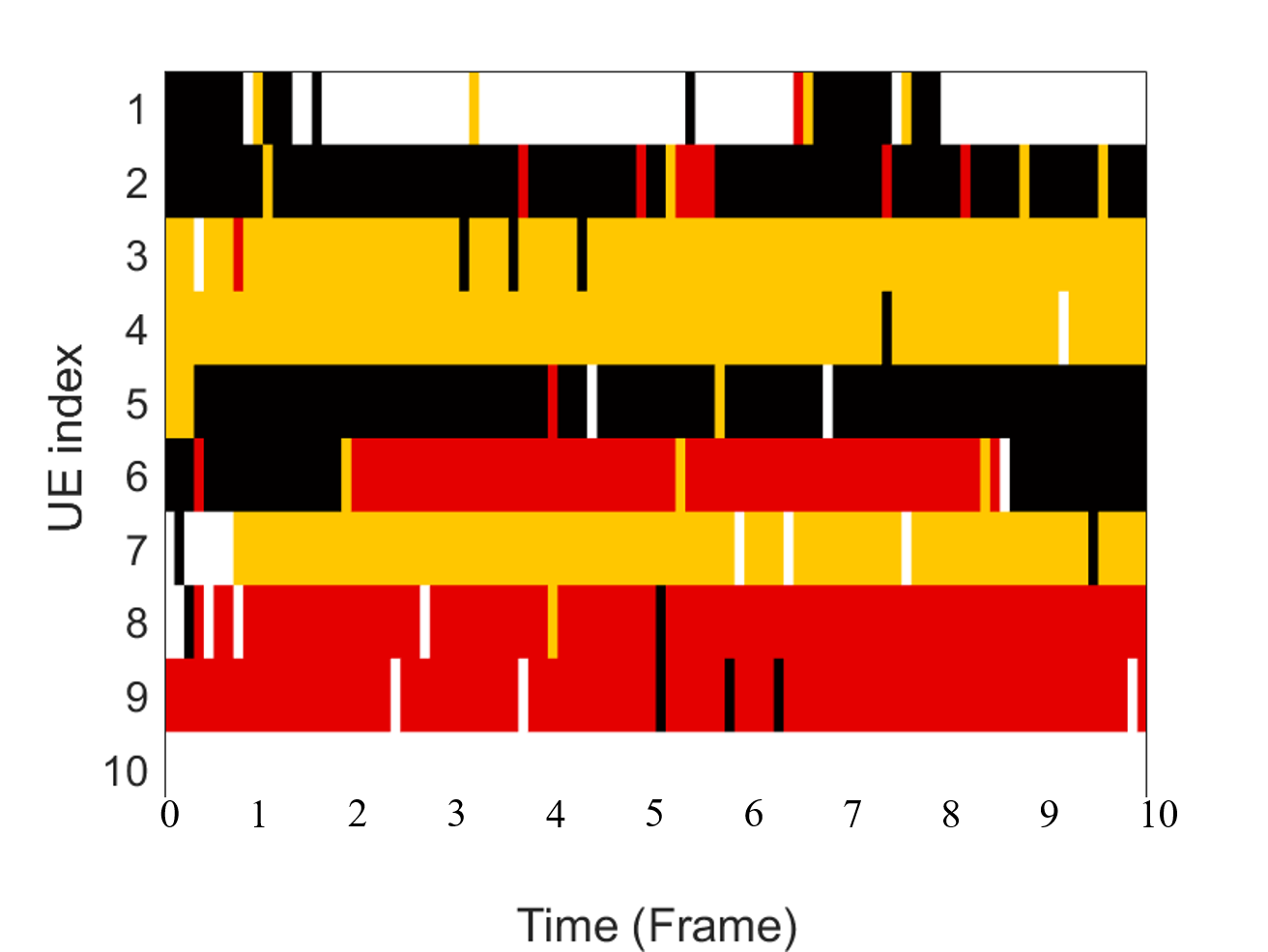}
    \caption{The channel selection using MARL method over time, where there are 10 UEs sharing 3 channels. The yellow, red and black represent the three channels and the white represents the case that the UE stays silent.}
    \label{fig: RL}
\end{figure}

We consider the competitive model where a UE's reward is based on its own throughput within a time slot. When there are sufficient channels for each UE, UEs are able to share the available channels with few collisions as claimed in~\cite{naparstek2018deep}. However, when there are more UEs than channels, directly applying MARL leads to two issues as shown in Fig. \ref{fig: RL}.

First, the majority of the UEs choose to request channels, leading to serious collisions all the time and none of the UEs can successfully communicate. The reason is that all the UEs are so selfish that they prefer to request channels even with a slight probability to obtain access when the other competitors turn to random choice than staying silent.\footnote{In reinforcement learning, we take the $\epsilon$-greedy policy, where agents take the random choices with little probability $\epsilon$.} A more profound explanation of these collisions by Nash Equilibrium is provided in Sec \ref{sec: design_MARL}.

Second, UEs tend to either occupy a particular channel or never request channel over time except the $\epsilon$-greedy policy. Such behavior is also harmful even if the collision does not occur. In this case, some UEs can always occupy channels and transmit data while others can never communicating data through these channels, hence, the fairness issue arises. The reason is that the action history that UEs store is limited by several factors, including UEs' memory size and the input size of DQN model.
With the limited memory size, it is impossible for UE to learn an alternative choice pattern. As an extreme result, UEs stick to particular channels all the time.

\subsection{Observation and Motivation}\label{sec: back_observation}
We implement the above MARL method under different settings and analyze the trained models of each UE. We make two key observations from the existing MARL methods.

\begin{figure*}[ht]
    \centering
    \includegraphics[width=1.8\columnwidth]{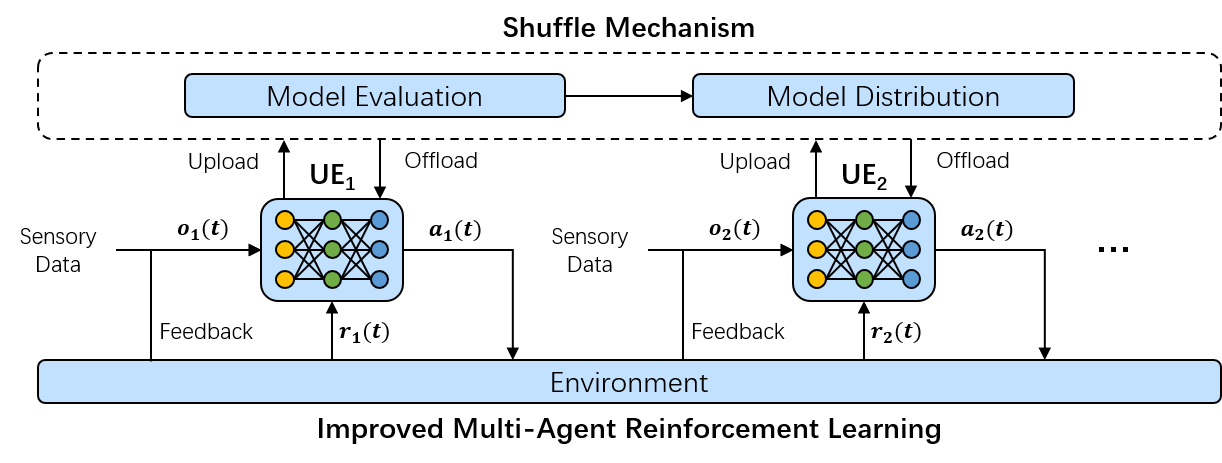}
    \caption{A closer look at the Hermes architecture.}
    \label{fig: diagram}
\end{figure*}

First, the longer the models are trained and the closer to the convergence, the more diverged the models across UEs are. Specifically, not only the models that converge to the different channels have different parameters, but also the models that prefer to choose the same channel diverge. 
Such an observation indicates that these models do not need to be unified and the discrepancy among converged models should be maintained.

Furthermore, the aforementioned model divergence does not result from the input sensory data. Instead, the subtle difference of random initialization of models leads to the significant discrepancy of the converged models. The interaction across UEs is the only feasible way to train a model that converges to choose a particular channel. This observation suggests that we cannot modify the models' parameters, but either reuse an existing model or train models from the scratch.

Based on those two key observations, we propose a novel shuffle mechanism in Hermes, where we shuffle the uploaded models and distribute them back to UEs. Such a shuffle mechanism maintains the diversity of models and reuses all the existing models for the next period. As a result, after shuffling the models, UEs are likely to change the channel choice, e.g., from staying silent to requesting a particular channel and vice versa. Therefore, the fairness issue can be effectively addressed.

\section{Framework Design}\label{sec:design}
\subsection{Overview}\label{sec:overview}
Fig. \ref{fig: diagram} shows a closer look at the Hermes architecture. As shown, Hermes consists of two modules: the improved MARL (iMARL) framework and the shuffle mechanism.

In iMARL, we deploy a DQN on each UE to learn the Q function. For every time slot, each UE inputs sensory data and feedback from the environment, which constitutes the observations, to the DQN and takes an action according to the DQN outputs. The environment collects the actions from all the UEs and provides rewards and feedback for UEs. Then the DQN will be updated based on the observations, actions and rewards.

After several updates of DQN, the local models from UEs will be uploaded to the shuffler, where the proposed shuffle mechanism is performed. We execute shuffle mechanism in two steps: \textit{model evaluation} and \textit{model distribution}.
First, the model evaluation module estimates the potential UE preference for the received models from UEs.
The model distribution module will then fairly distribute corresponding models back to UEs based on the estimated preference.

\begin{figure}[ht]
    \centering
    \includegraphics[width=0.95\columnwidth]{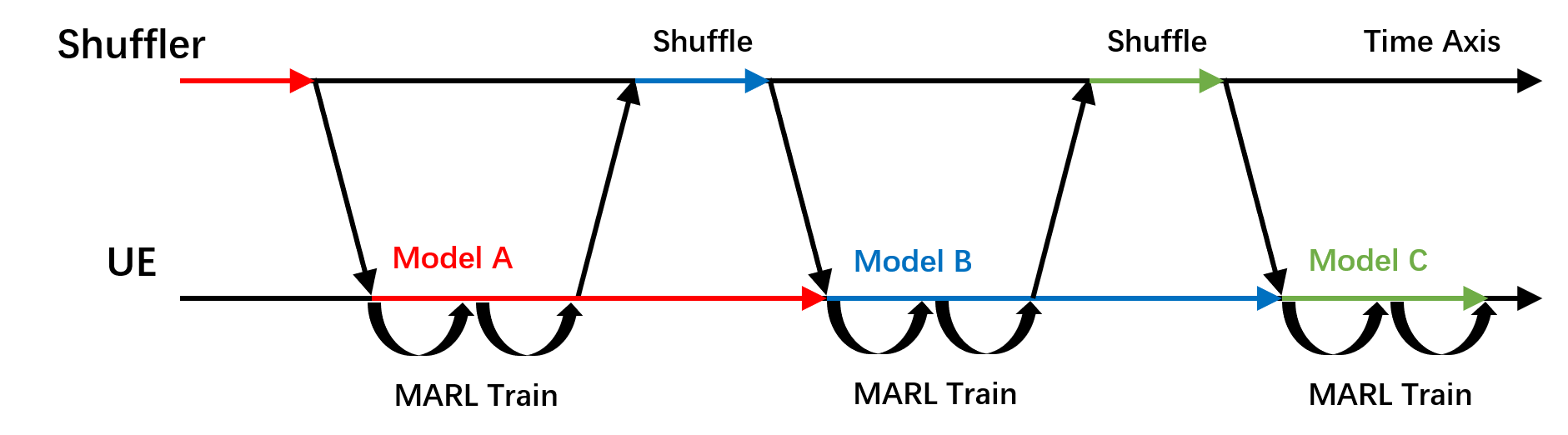}
    \caption{The work flow of Hermes.}
    \label{fig: timeline}
\end{figure}

Fig.\ref{fig: timeline} shows the workflow of Hermes. Once a DQN model is distributed by the shuffler, the UE keeps running and training this model until the next model arrives. It means that UEs never wait for the next shuffled model to come. The training of iMARL is only performed between two shuffling cycles. After UEs upload their models to the shuffler, they merely run the model, but not train the model any longer. Training is not taken in parallel with shuffles because the trained models will be overwritten by the incoming shuffled models.

Although training and shuffling the DQN model incurs certain computation cost, the critical bottleneck of the delay is only the inference latency by the DQN. Therefore, given a DQN with efficient inference, Hermes can offer very responsive spectrum management.

\subsection{Design Challenges}
\label{sec:challenges}
The combination of iMARL and shuffle mechanism in Hermes is promising for decentralized DSA, however, it also introduces three design challenges.

\noindent\textbf{Challenge 1.}
As described in Section \ref{sec:background}, one of the major drawbacks in the existing MARL based method is channel collision due to UEs' greediness. Considering the decentralized settings where UEs do not interact with each other, it is challenging to design iMARL that converges to a model that can maximally avoid collisions and significantly improve the channel utilization.

\noindent\textbf{Challenge 2.}
Traditional DQN architecture is heavy in two aspects: First, deep neural networks often contain a large number of parameters. Even worse, the Q network and the target Q network in DQN double the model size. Second, a replay buffer that stores previous experience is attached to each model. Sharing the DQN model between UEs and the shuffler incurs significant communication overhead on throughput. How to design a compact DQN structure to reduce communication overhead becomes a great challenge.

\noindent\textbf{Challenge 3.}
As presented in Section~\ref{sec: back_observation}, the DQN model are heterogeneous across UEs. The shuffler needs to match the DQN model with UEs based on their potential preferences for channel selection. However, it is difficult to deduce the channel selection from the received model parameters. Moreover, the sensory data cannot be explicitly shared with the shuffler. Hence, the preference evaluation can only be performed using the received models. Therefore, how to efficiently evaluate the personalized preference of each UE w.r.t each model without sensory data poses another challenge.

Addressing these challenges is not trivial. Without careful design of the DQN model, the communication overhead of sharing the DQN model can overshadow the benefits they bring. Even worse, if the model does not converge to a scenario with little collision, the channel cannot be efficiently utilized and is wasted. In addition, the evaluation module at the shuffler needs to be carefully designed, such that the UEs can be well matched to the model without raw sensory data. In the following, we present the techniques we develop in Hermes to effectively address those challenges.

\subsection{System Configurations}
Before diving into the design details, we need to clarify the system configurations of Hermes, which significantly impacts our design.
In terms of the wireless environment, we assume there are $N$ UEs to share $M$ channels, where $N>M$. The achievable bits transmitted per unit of time, or termed as data rate, offered by each channel varies among different UEs and can be estimated by CQI. 

In one time slot, a UE can either stay silent or request one channel to transmit data. A silent UE does not send any data or interfere with any channel. If a UE requests a particular channel, it can successfully transmit data when the channel is only requested by this UE. Otherwise, when two or more UEs request the same channel, a collision occurs and all UEs fail to transmit. At the end of this slot, UEs obtain feedback from the environment. The feedback is about whether they have transmitted data successfully or not. However, UEs cannot know peers' choices for the channels and which UE has a collision with them.

In addition, each UE has a local buffer that temporarily stores the data to transmit. In our settings, we only care about the amount of buffered data. The amount will increase when the UE has new data to transmit but does not get sufficient channel resources instantly. In contrast, when a UE is able to transmit data through a certain channel, the amount of buffered data will decrease accordingly. The data transmission will be terminated once the buffer is empty but the channel will be continuously occupied in the rest time of the current slot.

\subsection{Design of the Improved Multi-Agent Reinforcement Learning}
\label{sec: design_MARL}

\begin{figure}[t]
    \centering
    \includegraphics[width=1\columnwidth]{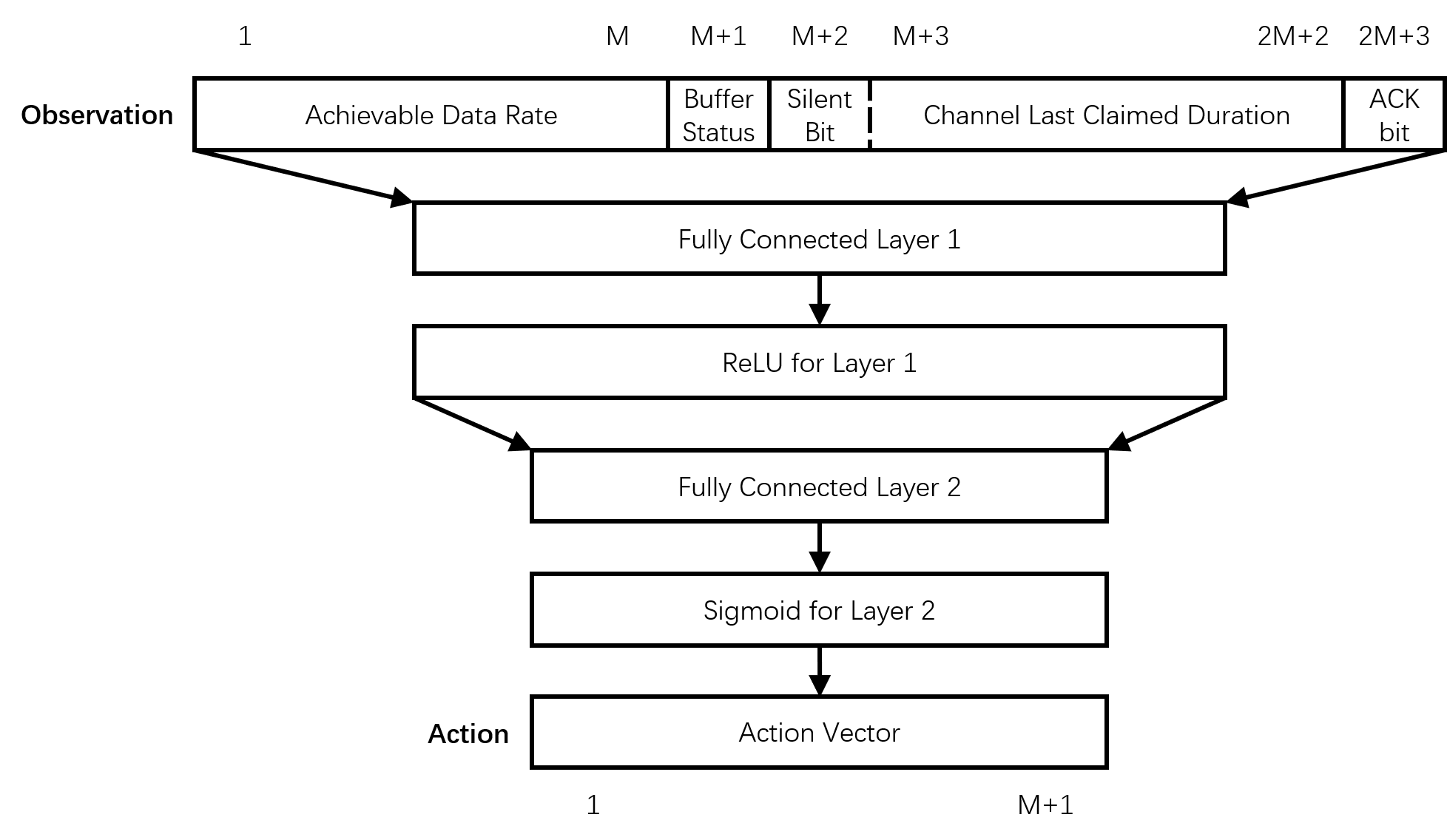}
    \caption{The structure of 2-layer-FC DQN.}
    \label{fig: dqn}
\end{figure}

We implement iMARL based on Deep Q Network (DQN)\cite{mnih2015human}. The structure of DQN deployed on each UE is shown in Fig. \ref{fig: dqn}. In this part, we will first introduce the core components of iMARL, including \textit{observations}, \textit{actions} and \textit{rewards}. Then, the compact DQN structure is presented. Finally, we will explain how we modify the DQN's workflow to further adapt the shuffle mechanism.

The observation $o_i(t)$ of the $i$th UE at time slot $t$ consists of four parts, {\color{black}whose space is

\begin{equation}
    o_i(t) \in \mathbb{R}_+^M \times \mathbb{N}_+ \times (\{0, 1\} \times \mathbb{N}_+^M) \times \{0, 1\}
\end{equation}

\noindent
where $\mathbb{N}_+$ and $\mathbb{R}_+$ represent the set of positive integers and positive real numbers.
}
The first $M$ elements are the available data rate offered in $M$ channels. This part is the sensory data that is collected by the UE's sensor and is independent of the actions. The second part is the buffer status. It has only one bit indicating whether the buffer is empty or not. The next $M+1$ elements record the UE's action history in the previous slots: the first element indicates whether the UE stays silent or not in the last slot; and the following $M$ elements record how long from the last request for the $M$ channels accordingly. The last element refers to whether the UE has successfully transmitted data in the last slot.

{\color{black} Let $a_i(t)$ denote $i$th US's action at time slot $t$, which can be formulated as:

\begin{equation}
    a_i(t) \in \{1, 2, 3, ..., M, M+1\},
\end{equation}

\noindent
where the former $M$ elements represent the preference for choosing each one channel from the $M$ channels and the last element represents the preference for staying silent.
}
For the time slot $t$, the $i$th UE's action $a_i(t)$ is determined by the $\epsilon$-greedy policy.
After an inference of DQN, the output is considered as the action vector. {\color{black}The action vector contains $M+1$ elements, each of which is corresponding to a choice of $a_i(t)$.} In $\epsilon$-greedy policy, UEs choose the best action with the maximal value in the action vector with probability $1-\epsilon$ and take random choice in small probability $\epsilon$.

\begin{figure}[ht]
	\begin{minipage}[t]{1\linewidth}
	    \centering
	    \subfloat[Reward Table of Existing MARL]
		{\includegraphics[width=0.45\columnwidth]{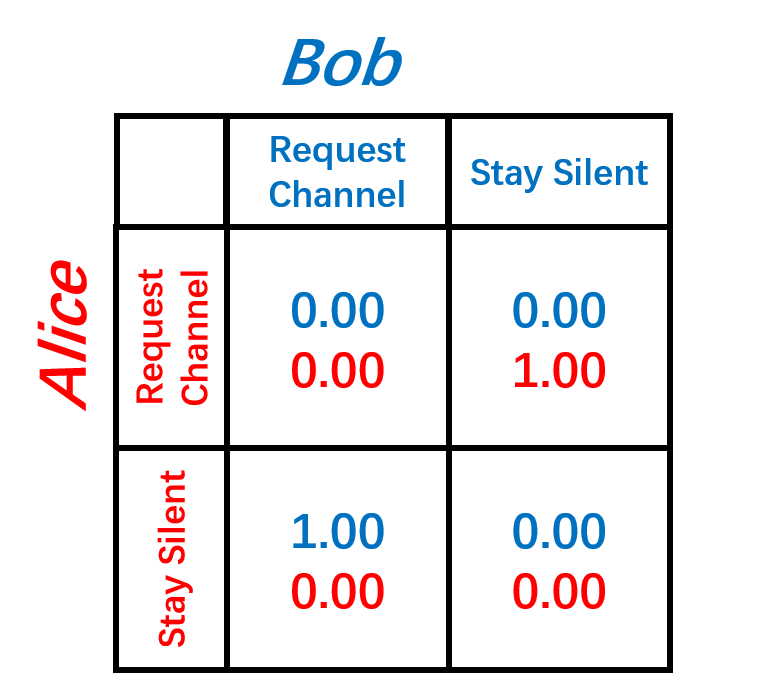}\label{fig: NashEquilibrium_1}}
		\subfloat[Reward Expectation Table of Existing MARL]
		{\includegraphics[width=0.45\columnwidth]{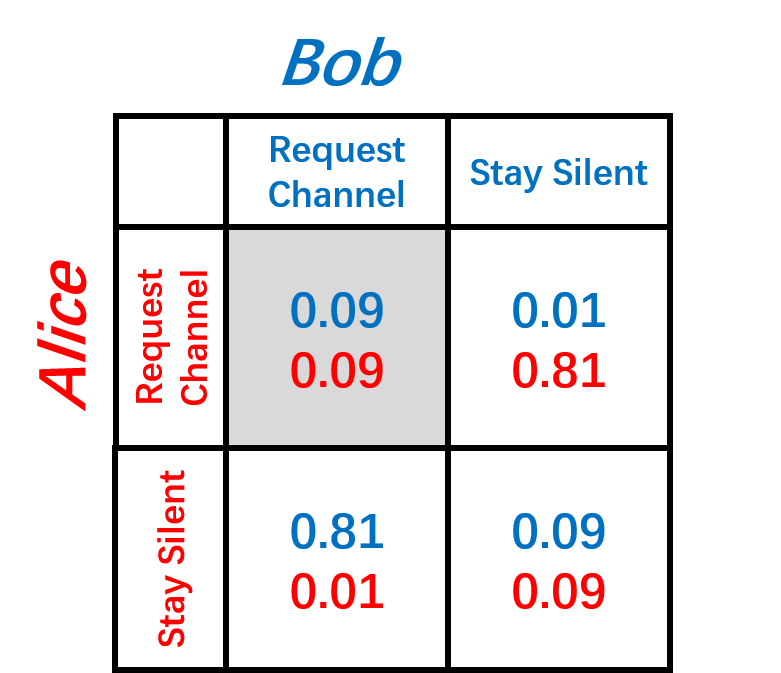}\label{fig: NashEquilibrium_2}}
		
		\subfloat[Reward Table of iMARL]
		{\includegraphics[width=0.45\columnwidth]{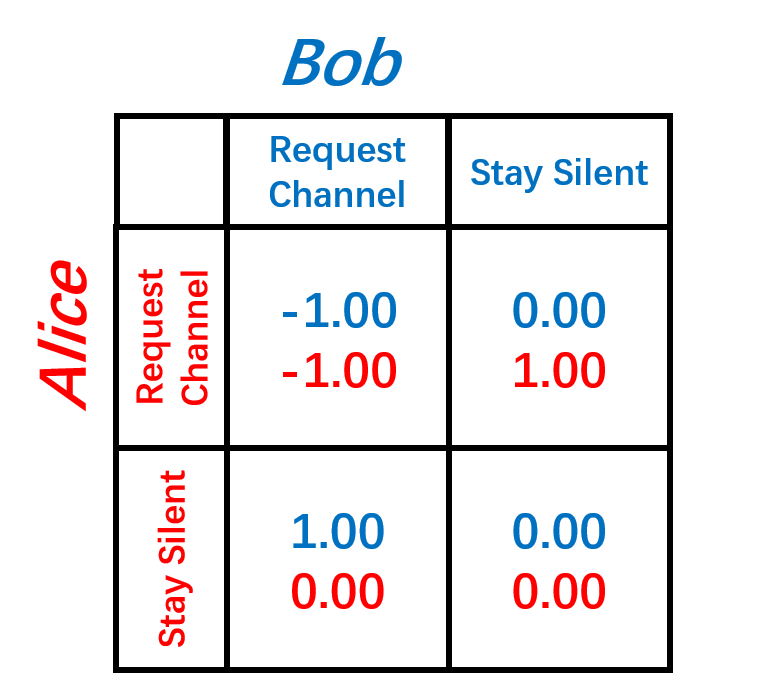}\label{fig: NashEquilibrium_3}}
		\subfloat[Reward Expectation Table of iMARL]
		{\includegraphics[width=0.45\columnwidth]{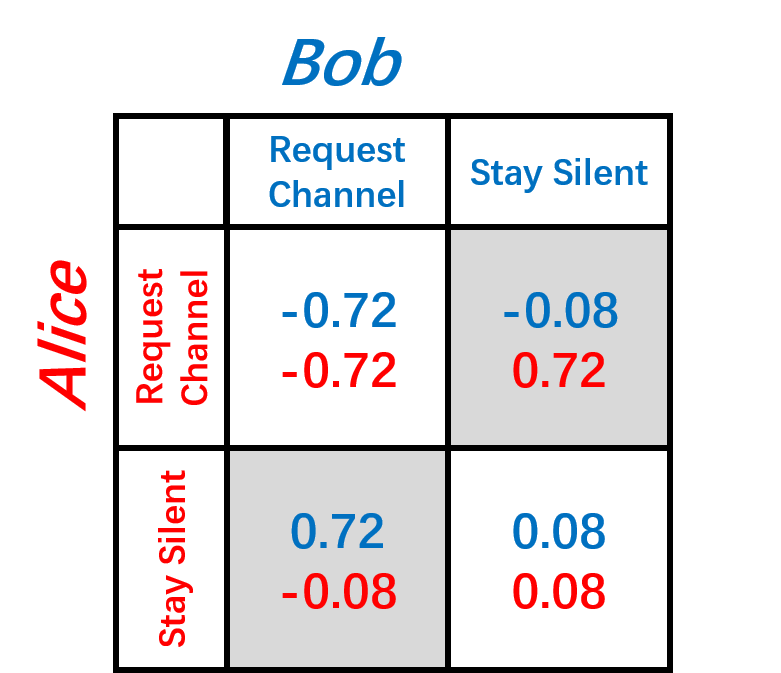}\label{fig: NashEquilibrium_4}}
		\caption{Reward tables and reward expectation tables of existing MARL and iMARL in a toy example, where two UEs, Alice and Bob, share one channel. The reward and reward expectation are painted as red and blue for Alice and Bob, and the gray scenarios are the Nash Equilibriums.}
		\label{fig: NashEquilibrium}
	\end{minipage}
\end{figure}

The setting of the rewards plays an important role in UEs' actions after convergence. Let $r_i(t)$ be the reward $i$th UE obtains at slot $t$.
In the existing MARL methods, $r_i(t)$ depends on its transmitted data, i.e., UEs get positive rewards when successfully transmitting data and zero rewards otherwise.
Such rewards lead to frequent occurrences of collisions, as demonstrated in a toy example in Fig.~\ref{fig: NashEquilibrium} where 2 UEs, Alice and Bob, share one channel.
The reward settings for Alice and Bob in existing MARL methods can be formulated as a reward table in Fig. \ref{fig: NashEquilibrium_1}. Because of the $\epsilon$-greedy policy, UEs do not always take the same actions even after convergence. Therefore, we cannot directly calculate the Nash Equilibrium from the reward table. Instead, a table containing the reward expectation over actions for each scenario is required. Let $\epsilon$ be $0.2$ in this toy example, which means there is $0.9$ probability to take the best action and $0.1$ probability to take the other action. Based on this probability distribution, we can calculate the reward expectation from reward table. In the scenario that both Bob and Alice request the channel, the possibilities of collision, Alice's using the channel alone, Bob's using the channel alone, and the channel's being idle equal 0.81, 0.09, 0.09, 0.01 respectively. We take the products between the possibilities and Alice's corresponding rewards in Fig. \ref{fig: NashEquilibrium_1} and sum the products up, yielding Alice's reward expectation as 0.09 in this scenario. Doing so for both Alice and Bob in all the four scenarios, we obtain the reward expectation table in Fig. \ref{fig: NashEquilibrium_2}.
In this reward expectation table, the scenario where both Alice and Bob request channel is the Nash Equilibrium. Hence, collisions occur frequently.
In contrast, negative rewards are assigned to the UEs with collisions in Hermes. As Fig. \ref{fig: NashEquilibrium_3} shows, the rewards of the collision scenario is set to negative values, $-1$ in this toy example, for both UEs. Similarly, the reward expectation table with $\epsilon$-greedy policy can be calculated as in Fig. \ref{fig: NashEquilibrium_4}. In this case, Nash Equilibrium moves to the scenarios where one UE transmits successfully and the other stays silent without collisions.
Generally, Nash Equilibrium is the situation where $M$ UEs stick to $M$ channels and $N-M$ UEs always stay silent in iMARL. Hereby, we define the reward function as:

\begin{equation}
    r_i(t)=\left\{
    \begin{array}{rcl}
    x_i(t) & & {\text{Success to transmit}}\\
    -\alpha * x_i(t) & & {\text{Fail to transmit}}\\
    0 & & {\text{Stay Silent}}
    \end{array} \right.
\end{equation}

\noindent
where $x_i(t)$ is the data rate and $\alpha$ is a punishment factor.
With the discounted factor $\gamma$, the total reward of the $i$th UE $R_i$ is defined as:

\begin{equation}
    R_i = \sum_{t=1}^{T}\gamma^{t-1}r_i(t).
\end{equation}

To reduce the overhead of sharing the DQN model, as well as the inference latency of DQN, the structure of DQN should be compact. Therefore, the DQN that we design for each UE contains two fully connected (FC) layers and two activation layers ReLU and Sigmoid respectively. The feedforward process of FC layers can be formulated as:

\begin{equation}
    \begin{split}
        x^{l+1}=&\ \mathbf{W^l}x^l+b^l\\
        (l=&\ 1,2)
    \end{split}
    \label{eq:FC layer}
\end{equation}

\noindent
where $\mathbf{W}$ is the weight matrix and $b$ is the bias vector; $x^l$ and $x^{l+1}$ are the input and output of the $l$th FC layer. We denote the size of the hidden layer as $L$ and the total parameters of the two FC layers in the network are $(3L+1)\times(M+1)$.

To train the DQN for each UE, we establish two identical networks, i.e., Q network and target Q network. The target Q network is fixed during the time that Q network is being trained. Once one training epoch ends, the Q network is copied to the target Q network. Also, a replay buffer is assigned to store the experiences and a batch of experiences are randomly sampled from the space in the training process. In the replay buffer, the tuples $[s_i(t), a_i(t), r_i(t), s_i(t+1)]$ at previous time slots are stored. The notation of the tuple is simplified as $[s, a, r, s']$ in the rest of the paper. To train the network, we define the loss function as follows:

\begin{equation}
    \begin{split}
    L(\theta)=
    E_{(s, a, r, s')}\Big[\big(r+\gamma \max_{a'} & Q(a',s';\theta^{target})\\
    -& Q(a,s;\theta)\big)^2\Big],
    \end{split}
\end{equation}

\noindent
where $\theta$ is the parameters of the Q network and $\theta^{target}$ is the parameters of the target Q network. It is easy to calculate the gradients of the loss function as in Eq.~\ref{eq: Loss Function Gradient} and the Q network is optimized with the expectations of gradients being estimated via sampled tuples in the batches.

\begin{equation}
    \begin{split}
    \nabla_{\theta}L(\theta)=
    & E_{(s, a, r, s')}\Big[ - 2\big(r+\gamma\max_{a'}Q(a',s';\theta^{target})\\
    & -Q(a,s;\theta)\big)\nabla_{\theta}Q(a,s;\theta)\Big].
    \end{split}
    \label{eq: Loss Function Gradient}
\end{equation}

To further reduce the overhead of communicating iMARL between UEs and the shuffler, two additional modifications are introduced. UEs only upload the DQN model right after one training epoch ends. At this moment, the Q network is about to be copied to the target Q network, so the two are supposed to be the same. Thus, only the Q network is uploaded. Also, since training no longer proceeds after uploading and the target Q network is never used for inference, the local copy can be omitted. The second modification is that the replay buffer is not uploaded and it is cleaned when a new shuffled model arrives. UEs train the DQN using data that is collected recently and locally. The rationale is that when the training is close to converging, the variance of the experience is low, so a small batch of experiences are able to provide an accurate estimation of the gradient expectation in Eq. \ref{eq: Loss Function Gradient}. Before being converged, the model changes rapidly and the memory that is generated a long time ago is out of date.

\subsection{Design of Shuffle Mechanism}

Shuffle mechanism contains two steps, model evaluation followed by model distribution. In the following, we will illustrate the design details of both steps.

In the model evaluation, we evaluate all the uploaded models by calculating the UEs' potential preference to them.
We denote the model form the $i$th UE as $i$th model. A preference matrix $E$ is introduced, where element $E_{i,j}$ represents the $i$th UE's preference to the $j$th model. For each $E_{i,j}$, we consider a combination of two metrics, namely, the model last appearance $E_{i,j}^{MLA}$ and the model distance $E_{i,j}^{MD}$.
The model last appearance (MLA) $E_{i,j}^{MLA}$ refers to how long from the the last allocation of $j$th model to the $i$th UE. The maximization of MLA means that UEs prefer models that are not allocated recently, increasing the models' mobility over UEs. As the mobility increases, UEs can access all the models and thus all the channels. This metric directly contributes to the fairness issue.

In spite of the above benefit, it is not appropriate to evaluate the UE's preference using MLA only. The reason is that MLA does not utilize the parameters of the models and ignores the UEs' adaptations to different models. 

Model distance (MD) $E_{i,j}^{MD}$ describes the discrepancy between the $i$th model uploaded from the $i$th UE and the $j$th model received from the $j$th UE. We tend to distribute a UE with the model that is similar to the one uploaded by this UE via minimizing the MD. The rationale is that UEs that are allocated with similar models are able to adapt the new models with fewer time slots, reducing unexpected outputs that may lead to collisions.

Before calculating MDs, the Model normalization is performed. In our DQN with two FC layers, we only consider the MD of the last FC layer that converts the features to the action vector.
As presented in Eq. \ref{eq:FC layer}, a UE takes the action with the maximum value in the action vector. This means we only care about the relative value and deducting the same row vector from different rows of $W$ and $b$ will not affect the UE's decision. Therefore, a zero-mean normalization for each column of $W$ and $b$ is performed before comparing the models. The normalization can be defined as: 

\begin{align}
    \hat{\mathbf{W}} &= \mathbf{W} - \frac{1}{N_{row}} \left[\begin{array}{ccc}
        1&\cdots&1  \\
        \vdots&\ddots&\vdots \\
        1&\cdots &1
    \end{array}\right]\mathbf{W},\\
    \hat{\mathbf{b}} &= \mathbf{b} - \frac{1}{N_{row}} \left[\begin{array}{ccc}
        1&\cdots&1  \\
        \vdots&\ddots&\vdots \\
        1&\cdots &1
    \end{array}\right]\mathbf{b}.
\end{align}

Based on the normalized $\hat{W}$ and $\hat{b}$, we can define MD by the Euclidean distance of $W$ and $b$ between two models as:

\begin{equation}
    E_{i,j}^{MD} = ||\hat{W_i} - \hat{W_j}||^2 + ||\hat{b_i} - \hat{b_j}||^2.
\end{equation}

Finally, we calculate the preference $E_{i,j}$ using the two metrics $E_{i,j}^{MLA}$ and $E_{i,j}^{MD}$ as:

\begin{equation}
    E_{i,j} = E_{i,j}^{MLA} - \lambda E_{i,j}^{MD},
\end{equation}

\noindent
where $\lambda$ is a hyper-parameter that balances the weights of the two metrics.

\begin{algorithm}[t]
\caption{Optimized UE-Model Matching Algorithm}
\label{alg: matching}
{\bf Input: $E$: Preference Matrix}\\
{\bf Output: $\mathcal{M}_{opt}$: The Optimal Perfect Matching}\\
\begin {algorithmic}[1]
\STATE {$Right = max(E), Left = min(E)$}
\WHILE{$Right > Left$}
    \STATE {$E_{th} = (Right + Left) / 2$}
    \STATE {Set $\Tilde{E}$ as empty}
    \FOR{$i$ in UE set}
        \FOR{$j$ in model set}
            \IF{$E_{i,j} > E_{th}$}
                \STATE{Append $E_{i,j}$ to $\Tilde{E}$}
            \ENDIF
        \ENDFOR
    \ENDFOR
    \STATE {$\mathcal{M} = Hungarian(\Tilde{E})$}
    \IF{$\mathcal{M}$ is a perfect matching}
        \STATE{$Left = E_{th}$}
    \ELSE
        \STATE{$Right = E_{th}$}
    \ENDIF
\ENDWHILE
\RETURN $\mathcal{M}$ as $\mathcal{M}_{opt}$
\end{algorithmic}
\end{algorithm}

Based on the estimated UEs' preference to models, the shuffler distribute models back to UEs. The model distribution can be formulated as a weighted bipartite perfect matching \footnote{In bipartite matching problem, perfect matching $\mathcal{M}$ is a matching that matches all vertices of the graph} problem. The $N$ UEs and the $N$ models are the two bipartitions and the preference matrix $E$ is the edge weight matrix between these two bipartitions. The Hungarian algorithm~\cite{kuhn1955hungarian} and Kuhn-Munkres (KM) algorithm~\cite{munkres1957algorithms} can be utilized to find a perfect matching $\mathcal{M}_{KM}$ which maximizes the summation of the matched weights:

\begin{equation}
    \mathcal{M}_{KM} = \arg\max_{\mathcal{M}}(\sum_{E_{i,j}\in \mathcal{M}}E_{i,j}).
\end{equation}

However, KM algorithm does not take fairness into account. In the matching with the maximized summation of weights, the extreme matches with much smaller weights than others cannot be excluded. To eliminate such circumstances, we propose an optimized UE-model matching algorithm. Different from KM algorithm, our matching algorithm aims at maximizing a threshold $E_{th}$. All weights within the matching are greater than or equal to the threshold. In other words, the minimum weight in the matching is maximized:

\begin{equation}
    \mathcal{M}_{opt} = \arg\max_{\mathcal{M}}(E_{th}) = \arg\max_{\mathcal{M}}\big(\min_{E_{i,j}\in \mathcal{M}}(E_{i,j})\big).
\end{equation}

The matching algorithm is presented in Alg. \ref{alg: matching}. The core idea of this matching algorithm is trial and error using a heuristic binary search.
We set the upper bound $Right$ and the lower bound $Left$ for $E_{th}$. Given the two trial bounds, the threshold $E_{th}$ is set as $ (Right+Left)/2$. A new unweighted subgraph is built based on the threshold $E_{th}$. An edge from the $i$th UE to the $j$th model is included in the subgraph if and only if $E_{i,j}$ is no less than the threshold $E_{th}$. Then, the Hungarian algorithm is performed. If the perfect matching exists, we reset the lower bound $Left$ as $E_{th}$. Otherwise, we reset upper bound $Right$ as $E_{th}$. We repeat this loop until the upper bound $Right$ and the lower bound $Left$ converges.
\section{Evaluation}\label{sec:evaluation}

\subsection{5G Simulation Setups}

We implement and evaluate Hermes in 5G network. We refer to the MATLAB 5G Toolbox \cite{MATLAB5GToolbox} for developing the 5G simulation environment.

\textbf{5G Frequency Domain and Time Domain.}
We first introduce the 5G's structure in the frequency domain and time domain.
In the frequency domain, the resource block group (RBG) is the smallest unit allocated to UEs.
Each RBG contains 16 resource blocks (RBs) which consist of 12 subcarriers {\color{black}and the subcarrier spacing is 15kHz.}
The CQI on each RB can be sensed by UEs and is represented by a positive integer. The CQI value is determined by the distance between the UE and the gNb and is not greater than $15$. It fluctuates randomly by $\pm 2$ every 200 ms.

As for the time domain, `frame' is the unit of simulation time, each of which lasts 10 ms.
Each frame consists of 10 slots, which are the smallest unit that can be allocated to UEs in the time domain.
UEs transmit 14 OFDM symbols within each slot. The first 2 symbols represent the demodulation reference signal (DMRS) for channel estimation and the rest 12 symbols express data to transmit.

To calculate UEs' current data rate in an RBG for a slot, the mean value of CQI over the 16 RBs in this RBG is calculated to determine the modulation and coding scheme (MCS). Then the bit transmitted per symbol per slot is given by the MCS.
Hence, the current transmitted data size in a RBG per slot can be formulated as:

\begin{equation}
\begin{split}
\#Bits = bit/symbol &* \big(\#(symbol) - \#(DMRS )\big) \\
&*\ \#(subcarrier) \;* \; \#(RB).
\end{split}
\end{equation}

\textbf{5G Application Configuration.}
The dynamic change of UEs' buffer size depends on the requirement of applications on them. We simplify the application configuration by making three assumptions as follows:
\begin{itemize}
    \item There is no extra control signals between the gNb and the UEs. All the throughput is data transmission.
    \item The host device of all the applications is UE, indicating that only uplinks from UEs to gNb exist.
    \item There is no need for retransmission.
\end{itemize}

Two properties are being considered for each application: $packetInterval$ in slot and $packetSize$ in byte. Each UE's buffer increases by $packetSize$ bytes every $packetInterval$ slots for all the applications on this UE.

\subsection{Evaluation Metrics}\label{Sec: 4.2 Metrics}

\textbf{Channel Utilization Efficiency.}
We adopt \textit{channel utilization efficiency} (CUE) to evaluate Hermes.
The traditional CUE refers to the rate of utilized data rate over the maximum data rate of a channel. It is not available in Hermes because a channel's maximum data rate varies from different UEs.
Therefore, in our experiments, we define CUE as the proportion of used RBGs over all available RBGs.
A lower CUE indicates that there exist collisions or idle channels.
In other words, CUE can also be defined as one minus the proportion of collisions and idle RBGs:
\begin{equation}
    P(utilized) = 1 - P(collided) - P(idle).
\end{equation}

\textbf{Average Throughput.}
In the wireless network, throughput is an important metric that evaluates the data rate of wireless devices. We use \textit{average throughput} over UEs as one evaluation metric in our experiments.
With a fixed number of UEs, the average throughput over UEs is linearly proportional to the total throughput. Unlike CUE, average throughput measures how much capacity of RBGs are utilized, instead of merely counting the number of used RBGs. In this sense, average throughput can be viewed as the weighted CUE using RBGs' data rate.

\textbf{Jain's Fairness Index.}
We also apply \textit{Jain's fairness index} (JFI)~\cite{jain1999throughput} to evaluate the throughput fairness over UEs.
It is formulated as:
\begin{equation}
    \mathcal{J}(x_1, x_2, ..., x_N) = \frac{(\sum_{i=1}^{N}x_i)^2}{N\sum_{i=1}^{N}x_i^2},
\end{equation}
where $x_i$ is the throughput of the $i$th UE. As we can see in the equation, one advantage of this metric is that it is only impacted by the relative throughput of different UEs, but not the absolute values. As the system is the most fair and the throughput is even for all users, JFI reaches its maximum value of 1.

\subsection{Baselines}\label{sec:base methods}

We evaluate Hermes's performance by comparing it with the centralized PF method~\cite{tiwari2014long} and the decentralized DQSA method~\cite{naparstek2018deep}.

PF is a traditional scheduling strategy on gNbs in SM. Compared to other scheduling strategies such as Best CQI and Round Robin, PF is known for its better performance on fairness.
In the PF approach, UEs need to upload their full status to the gNb for every scheduling periodicity, including buffer status, CQI and historical average data rate. The historical average data rate describes the previous data rate for a UE and is updated every time slot.
For each RBG, PF tends to distribute it to the UE with the maximal proportion of the current data rate over the historical average data rate until a maximum RBG number (1 in our implementation) is allocated to this UE.

Besides, the decentralized method DQSA is taken as another baseline for Hermes. It is one of the state-of-the-art MARL based DSA method that outputs a good DSA scheduling plan in small-scale UE deployment. However, as described in Sec. \ref{sec:back_MARL}, DQSA suffers fairness issue and serious collisions when the number of UEs $N$ is greater than the number of RBGs $M$.

\subsection{Evaluation of Performance}
We first evaluate Hermes under a relatively simple setting, where only a small number of UEs and RBGs need to be scheduled in short simulation time. 
After that, we compare Hermes's with two baseline methods based on the aforementioned evaluation metrics under a more challenging setting compared to the above setting. In this setting, we increase the number of UEs and RBGs with the simulation time duration extended.
In both settings, we assume that sensory data across all the UEs are independent and identically distributed (i.i.d) and only one single shuffler serves all the UEs. The training of iMARL is performed every 10 slots and the shuffling is performed every 50 slots.

\begin{figure}[t]
	\begin{minipage}[t]{1\linewidth}
		\subfloat[Channel Selection]
		{\includegraphics[width=0.48\textwidth]{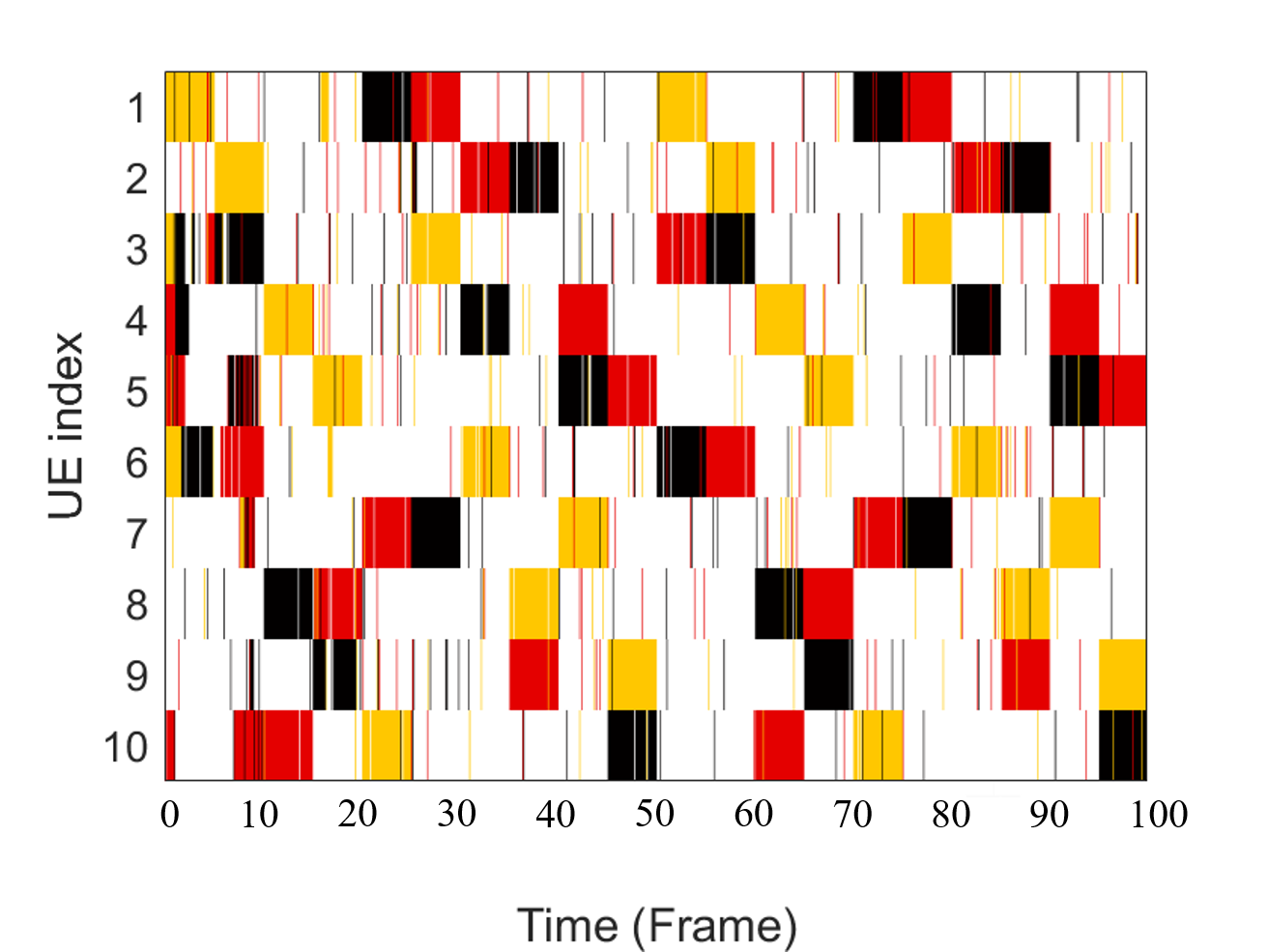}\label{fig: channel_selection_Hermes}}
		\subfloat[Throughput]
		{\includegraphics[width=0.48\textwidth]{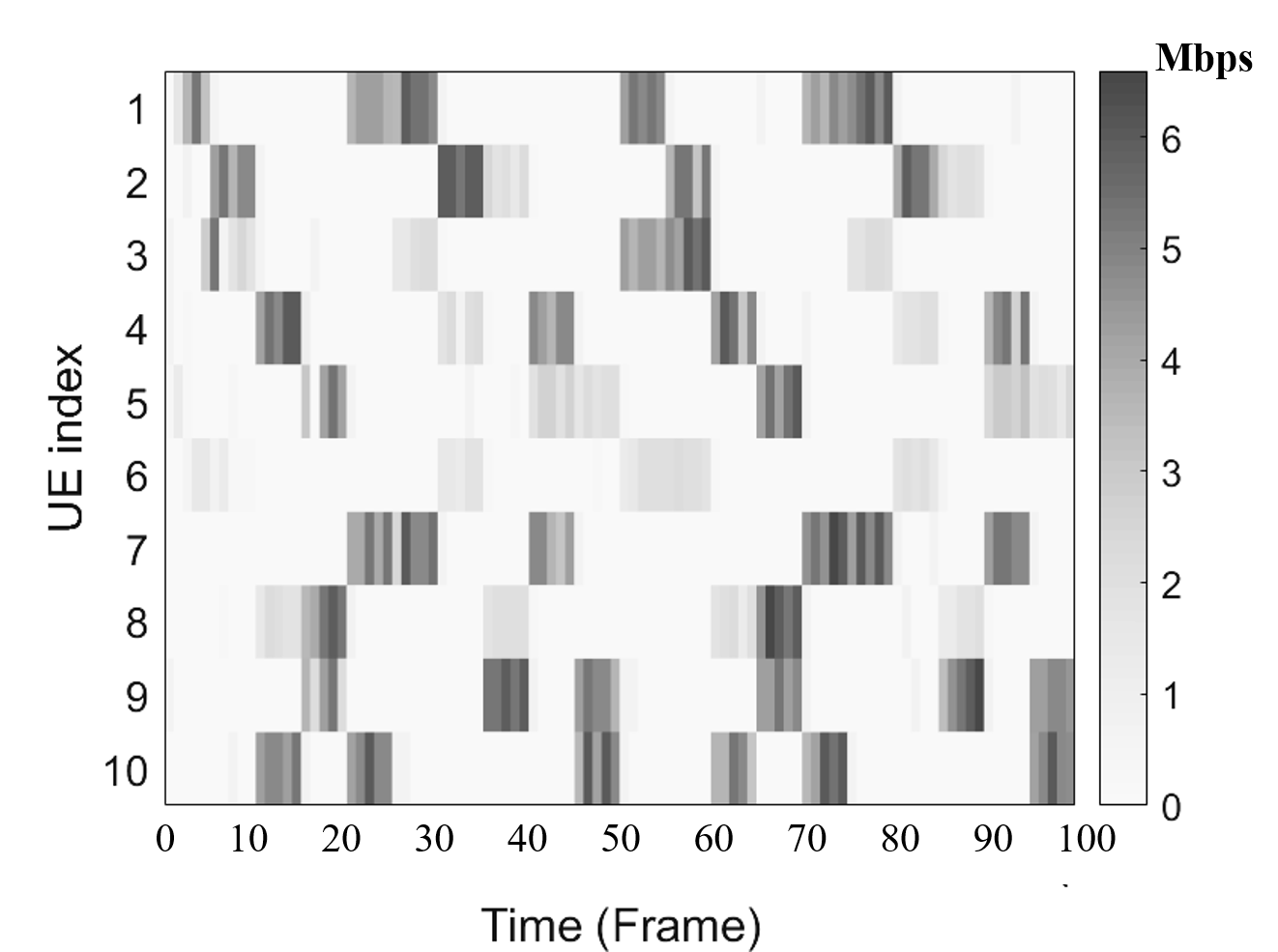}\label{fig: throughput_Hermes}}
		\caption{The channel selection and throughput heatmaps of Hermes, where there are 10 UEs sharing 3 RBGs for 100 frames.}
		\label{fig: timeline_Hermes}
	\end{minipage}
\end{figure}

Fig. \ref{fig: timeline_Hermes} shows the results under the relatively simple setting with 10 UEs and 3 RBGs for 100 frames (1 seconds or 1000 slots).
In Fig. \ref{fig: channel_selection_Hermes}, the color yellow, red and black represent the three channels and the color white represents that a UE stays silent. The 10 UEs start with random channel selections and end up with the Nash Equilibrium within 100 frames, where 3 of 10 UEs request channels and the rest of them stay silent. Note that the training of iMARL proceeds smoothly despite that shuffling takes place every 50 slots. Besides, there are still a small number of stripes after the DQN converges. It indicates that UEs retains a small tendency to explore different RBGs due to the $\epsilon$-greedy policy. Although such tendency leads to collided and idle channels, it helps the model to adapt to environmental changes. The model cannot make use of newly available RBGs any longer if the $\epsilon$-greedy policy is abandoned once after the DQN converges.

In terms of throughput, as Fig. \ref{fig: throughput_Hermes} shows, the darker the color stripe is, the higher the throughput UEs achieve. Generally, a channel can achieve a data rate of 3 Mbps to 6 Mbps. In the first 10 frames when models are not converged and collisions occur frequently, the throughput over all the UEs remains to a low extent. When the Nash Equilibrium is reached, the throughput heatmap looks similar to the channel selection heatmap, which indicates that almost all the requests for channels can successfully transmit data without collisions.

\begin{figure}[ht]
	\begin{minipage}[t]{1\linewidth}
		\subfloat[Channel Utilization Efficiency]
		{\includegraphics[width=0.48\textwidth]{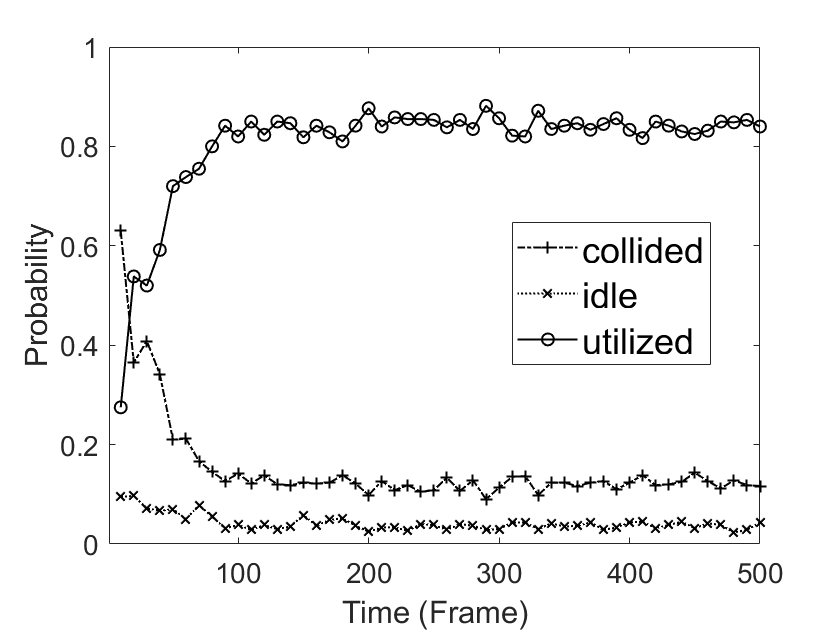}\label{fig: utilization efficiency}}
		\subfloat[Average Throughput]
		{\includegraphics[width=0.48\textwidth]{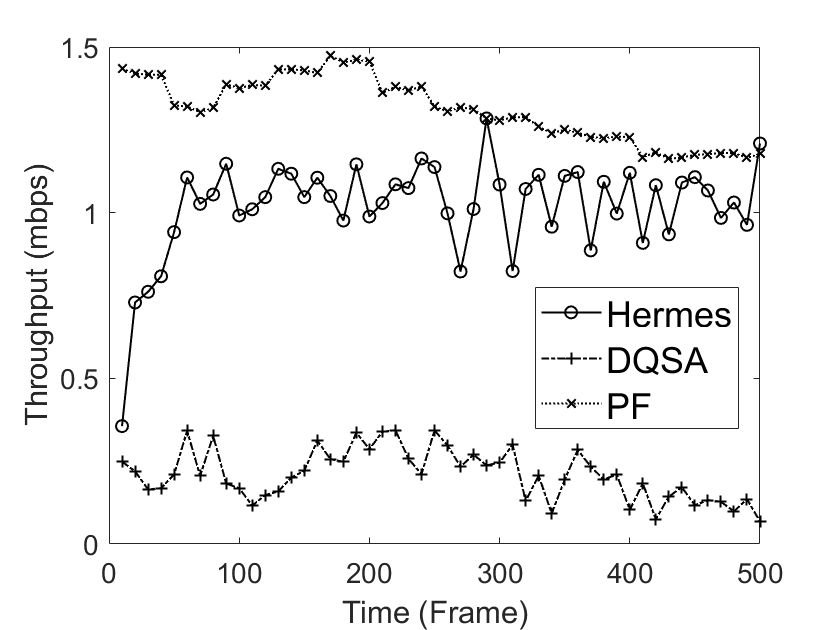}\label{fig: average throughput}}
		\caption{The CUE and average throughput over time, where there are 20 UEs sharing 6 RBGs for 500 frames.}
		\label{fig: 20 UE 6 RBG 5 seconds}
	\end{minipage}
\end{figure}

Then we compare Hermes with two baselines in a more challenging environment, where 20 UEs share 6 RBGs for 500 frames. Their performance is shown in Fig. \ref{fig: 20 UE 6 RBG 5 seconds} and Table \ref{tab: Jain's Fairness Index}.

Fig. \ref{fig: utilization efficiency} shows the CUE of Hermes. The DQN model converges around the 100th frame. After that, the CUE slightly fluctuates around the maximum and the collision probability is reduced to the minimal value. The probability of occurring idle channels keeps a small value all the time. The CUE cannot reach 100\% due to the $\epsilon$-greedy policy.

The average throughput of Hermes, DQSA and PF is shown in Fig. \ref{fig: average throughput}. The average throughput of Hermes is slightly lower than that of PF. The reason is that Hermes cannot fully utilize all available RBGs owing to the $\epsilon$-greedy policy. Yet, its average throughput is much higher than that of DQSA due to the properly designed reward.

We also evaluate JFI on all the three methods and report the evaluation results in Table \ref{tab: Jain's Fairness Index}. Hermes is slightly better than PF and these two methods are significantly better than the DQSA without shuffling in terms of fairness.

\begin{table}[t]
    \centering
    \caption{Comparison of JFI between Hermes, DQSA and PF.}
    \begin{tabular}{lc}
    \hline
        Methods & Jain's Fairness Index \\
        \hline
        Hermes & 0.9619 \\
        PF & 0.9274 \\
        DQSA & 0.4902 \\
        \hline
    \end{tabular}
    \label{tab: Jain's Fairness Index}
\end{table}

\subsection{Evaluation of Robustness}
In addition to performance evaluation, we evaluate Hermes's robustness with different numbers of UEs, RBGs and varied deployment intervals between UEs.
Since the average throughput is strongly correlated to the number of RBGs, the trend of the average throughput as RBGs increase tells little about the system's robustness with respect to the number of RBGs. Hence, we only use the other two metrics in this section.

\begin{figure}[ht]
	\begin{minipage}[t]{1\linewidth}
		\subfloat[impact of \#UE]
		{\includegraphics[width=0.48\textwidth]{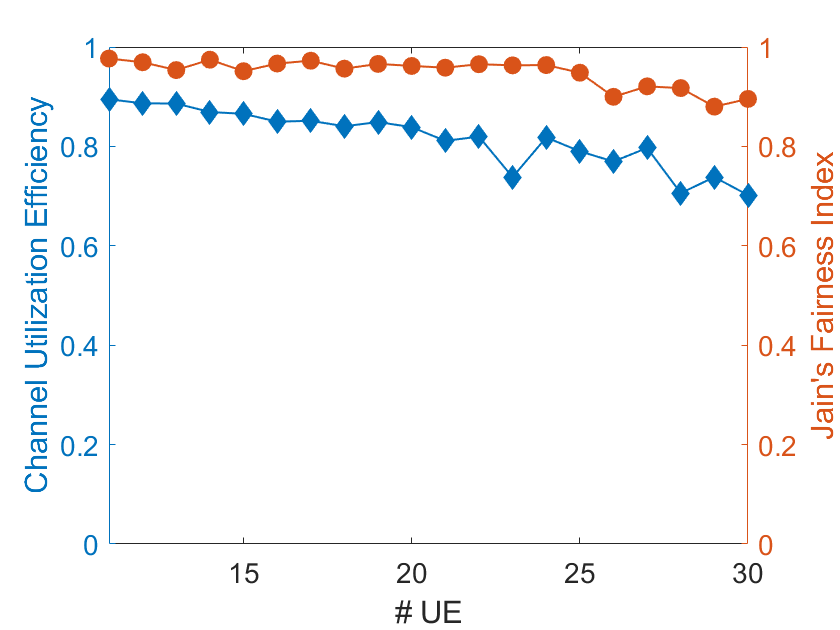}\label{fig: vsUE}}
		\subfloat[impact of \#RBG]
		{\includegraphics[width=0.48\textwidth]{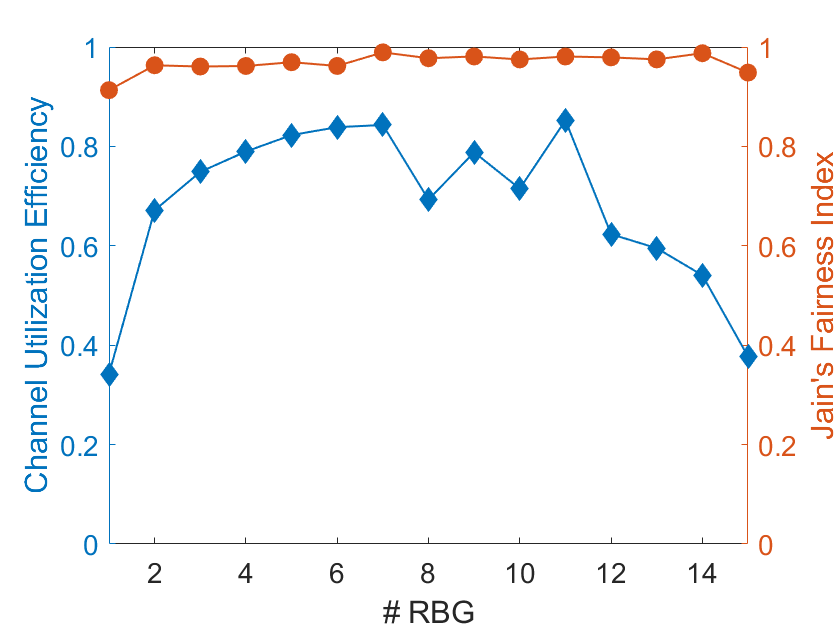}\label{fig: vsRBG}}
		\caption{The impact of \#UE numbers (a) and \#RBG (b) on CUE and JFI.}
		\label{fig: vsRBGnumUEnum}
	\end{minipage}
\end{figure}

We first evaluate Hermes's robustness with a varied number of UEs and RBGs. In this experiment, the simulation is conducted for 500 frames and we calculate the corresponding evaluation metrics based on the last 100 frames.
In Fig. \ref{fig: vsUE}, we vary the number of UEs from 11 to 30 with 6 RBGs shared. The CUE slightly decreases with a larger number of UEs. It is because the probability of colliding with other UEs when taking random choices increases, as we include more UEs.
Then, we vary the number of RBGs from 1 to 15 with a fixed 20 UEs. As Fig. \ref{fig: vsRBG} presents, Hermes achieves higher CUE when the number of RBGs is between 2 to 11,  but there is a significant performance drop in extremely competitive situations where the number of RBGs is too small. In addition, performance is also degraded when there are too many RBGs. The reason is that the DQN model converges so slowly that it has not converged at the end of 500 frames and hence the CUE remains low.
With regard to fairness, Hermes maintains a high JFI no matter how the number of UEs and RBGs are changed.

\begin{figure}[ht]
	\begin{minipage}[t]{1\linewidth}
		\subfloat[deployment]
		{\includegraphics[width=0.4\textwidth]{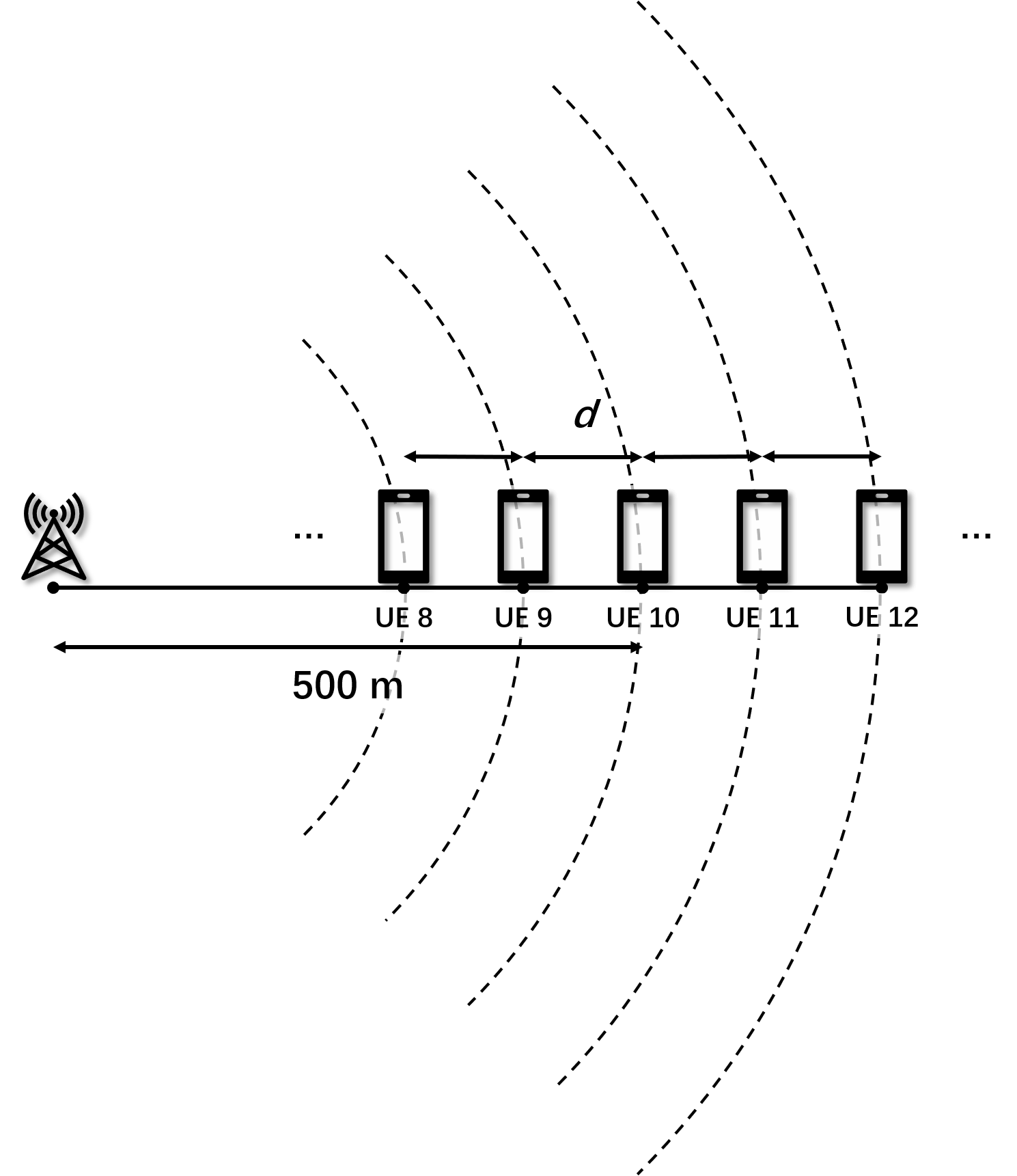}\label{fig: vsDist_deploy}}
		\subfloat[deployment interval $d$]
		{\includegraphics[width=0.6\textwidth]{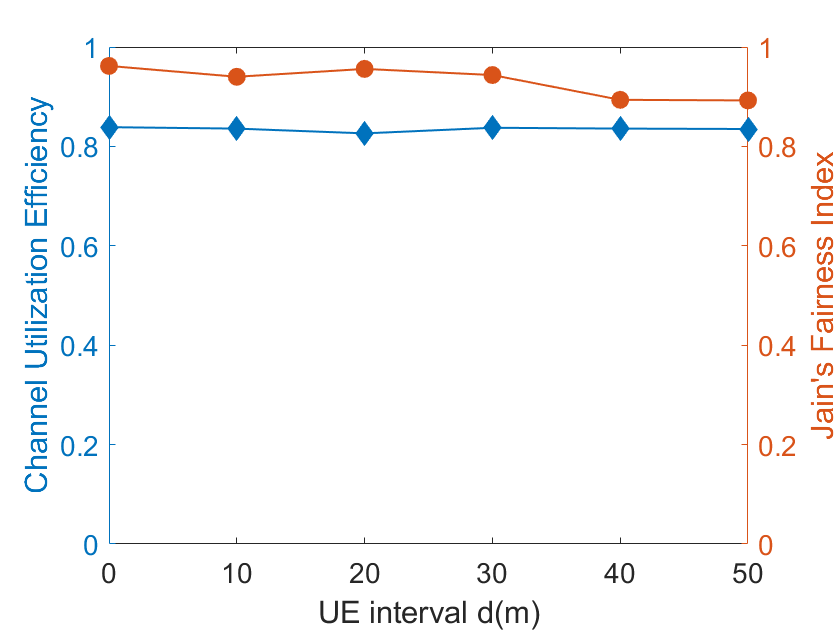}\label{fig: vsDist_metric}}
		\caption{The impact of deployment interval on CUE and JFI.}
		\label{fig: vsDist}
	\end{minipage}
\end{figure}

We also evaluate how the deployment interval between UEs affects the CUE and JFI. Here the deployment interval is defined as the distance between UEs. In this experiment, as shown in Fig. \ref{fig: vsDist_deploy}, we deploy all the UEs with the same interval $d$ in a line from the gNb, and the UE in the middle is always 500 meters away from the gNb. As Fig. \ref{fig: vsDist_metric} shows, the CUE is not impacted by the deployment interval. Meanwhile, the fairness can always be guaranteed even in the extreme situation, where the deployment interval $d$ is 50 meters. In that circumstance, the closest UE to the gNb is only 50 meters away and the farthest UE is at the edge of the gNb's coverage.

\subsection{Runtime Performance}
Finally, we evaluate Hermes runtime performance in a dynamic environment, where the numbers of available RBGs and UEs keep changing over time. In addition, multiple shufflers are deployed and each shuffler is responsible for shuffling the models among a random subset of the UEs. We apply the throughput to the evaluation of the runtime performance.

\begin{figure}[t]
	\begin{minipage}[t]{1\linewidth}
	    \centering
	    \subfloat[Total throughput]
		{\includegraphics[width=1\columnwidth]{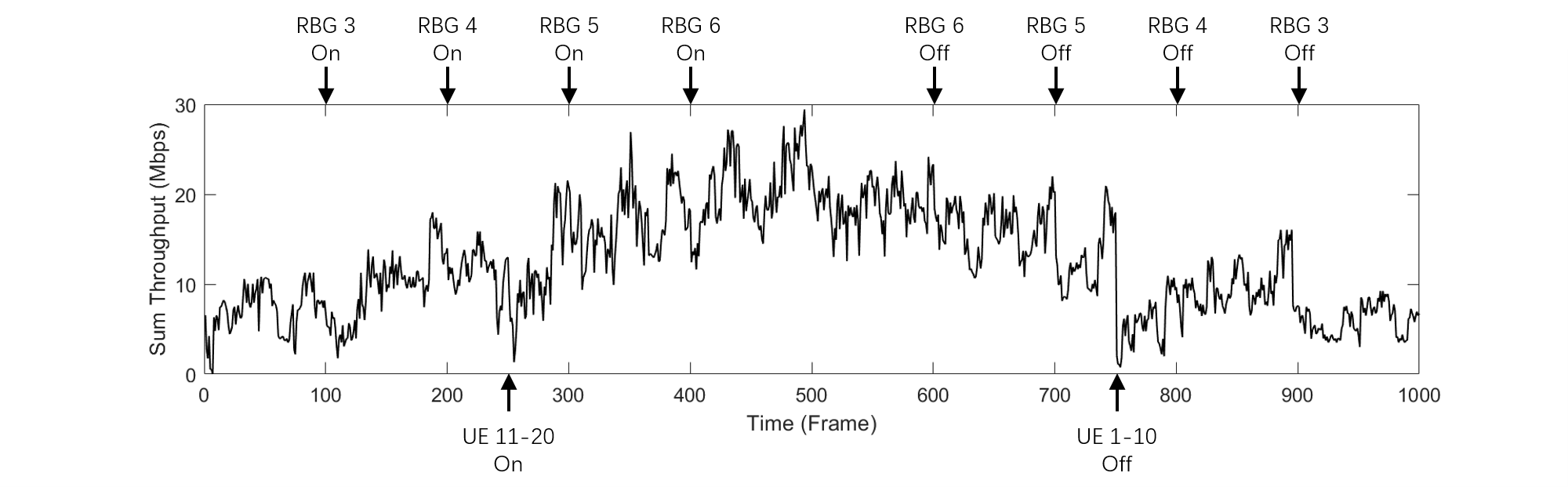}\label{fig: overallSum}}
	    
		\subfloat[Throughput over UEs]
		{\includegraphics[width=1\columnwidth]{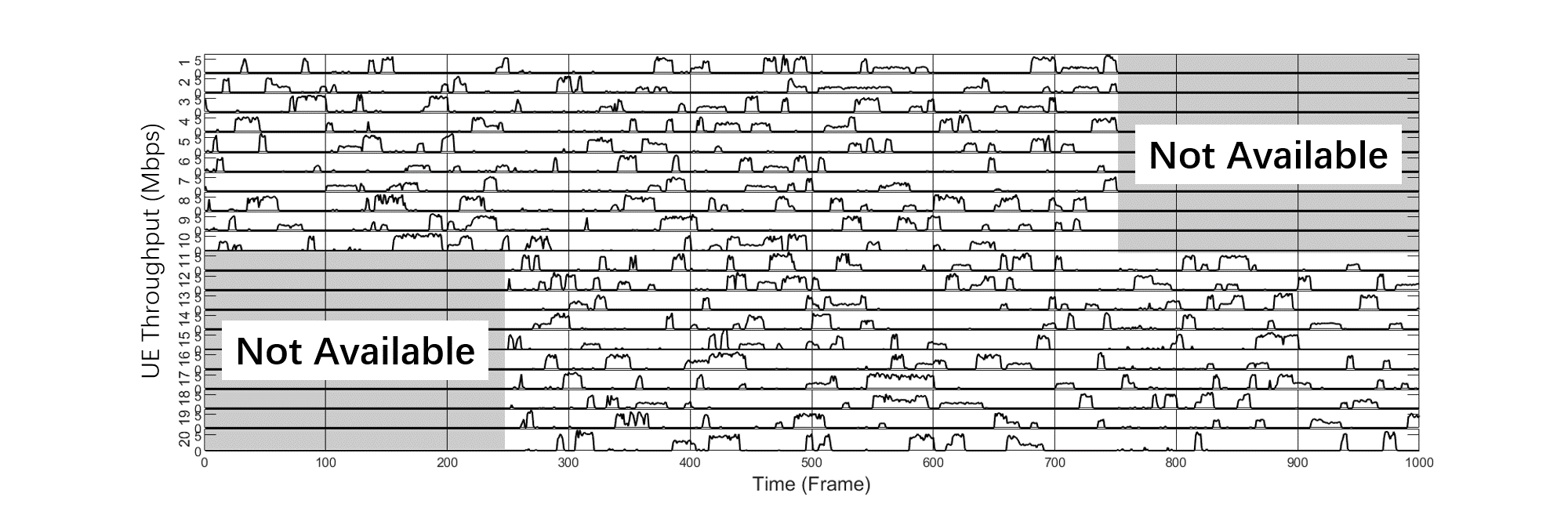}\label{fig: overallUE}}
		
		\subfloat[Throughput over RBGs]
		{\includegraphics[width=1\columnwidth]{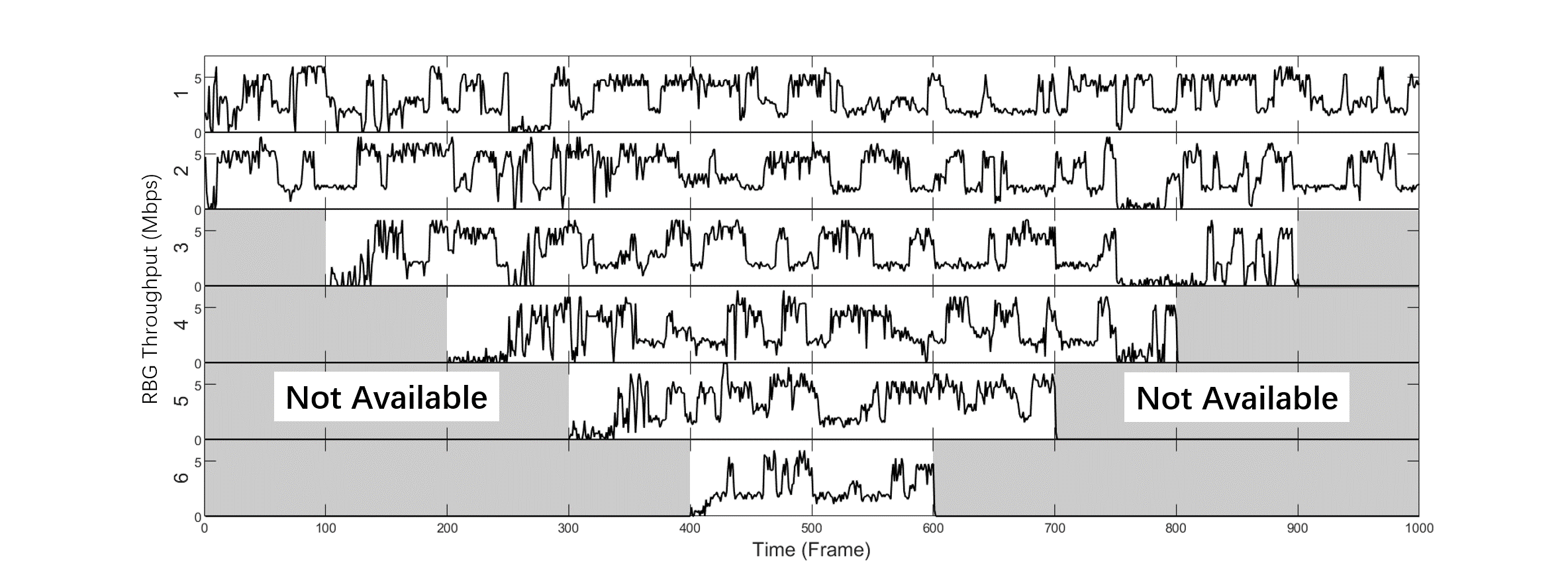}\label{fig: overallRBG}}
		\caption{The runtime performance of Hermes in the dynamic environment.}
		\label{fig: overall}
	\end{minipage}
\end{figure}

We simulate the dynamic environment for 1000 frames. There are 10 UEs and 2 RBGs at the beginning, and new UEs and RBGs are gradually added in the first 500 frames. Specifically, ten more UEs become available at the 250th frame and one more RBG is added every 100 frames from the 100th frame. In the next 500 frames, as shown in Fig. \ref{fig: overall}, some of the RBGs and UEs exit sequentially.

The results are presented in Fig. \ref{fig: overall}. When a new RBG becomes available, the throughput on this RBG increases slowly and reaches its full capacity in 0.5 seconds. On the contrary, when an RBG becomes unavailable, the corresponding throughput drops to zero immediately. The emergence of new UEs introduces a throughput drop within 0.5 seconds to all the UEs, because their untrained models bring collisions and are taken into the shuffle process and distributed to other UEs. When UEs exit, some of the RBGs become idle. The total throughput drops rapidly, but quickly recovers within around 0.5 seconds. To summarize, Hermes is capable of adapting to the environmental changes in no more than 0.5 seconds.
\section{Related Work}\label{sec:related_work}
The key challenge of designing a DSA protocol is to effectively and efficiently handle the multi-channel structure and highly dynamic network resources~\cite{akyildiz2006next}. Many DSA techniques have been proposed to address this challenge based on classical mathematical algorithms~\cite{arafat2017survey}. Among these techniques, the auction-based approach is a promising DSA technique. The auction-based techniques dynamically allocate the spectrum to SUs based on their best bid that is submitted to the PU. The bid can be expressed in many representations, e.g., money, relaying services, etc. For example, Wang \textit{et al.}~\cite{wang2009spectrum} designed a DSA approach based on bandwidth auctions, where SUs submit a bid for the spectrum and PUs allocate the spectrum among the SUs without affecting their own performance. Game theory is also adopted to optimize performance related to spectrum sharing. Maskery \textit{et al.}~\cite{maskery2009decentralized} proposed a DSA method from an adaptive game theory perspective. When a channel is available, SUs compete for it to fulfill their own demands while minimizing interference with their peers. Huang \textit{et al.}~\cite{huang2011cognitive} presented a non-cooperative game theory method for LTE-A networks, where femtocells compete for sharing the primary macrocell spectrum. On the contrary, Gharehshiran \textit{et al.}~\cite{gharehshiran2012collaborative} designed a cooperative game theory method for LTE-A networks. Niyato \textit{et al.}~\cite{niyato2008dynamics} proposed a comprehensive solution via combining cooperative and non-cooperative game theory approaches. Multiple PUs sell spectrum to multiple SUs, where the competition among PUs is modeled as a non-cooperative game and the competition among SUs is treated as an evolutionary game. Vamvakas \textit{et al.}~\cite{vamvakas2019dynamic} introduced prospect theory to allocate the transmission power as PUs or SUs. However, the dynamic interaction cannot be fully described using a game theory approach~\cite{saad2011coalitional}, and such an issue has been addressed by applying Markov Chains. Akbar and Tranter~\cite{akbar2007dynamic} proposed a Markov-based Channel Prediction Algorithm (MCPA), allowing SUs to dynamically select different licensed bands while significantly reducing the interference from and to PUs. Thao \textit{et al.}~\cite{nguyeny2011hidden} utilized the Hidden Markov Process to model the utilization state of each spectrum band at each time slot, and such a state can be approximated from the power spectral density measurements. Based on the occupancy state, the available bands are assigned accordingly.

Besides the mathematical approaches, RL has become an emerging state-of-the-art technique for DSA. The most widely applied RL algorithm in wireless communication applications is Q-learning~\cite{watkins1989learning}. Therefore, most of RL-based DSA approaches focus on Q-learning~\cite{morozs2015distributed,chen2012stochastic,nie1999q}. Morozs \textit{et al.}~\cite{morozs2015distributed,morozs2015heuristically} presented a centralized RL method for DSA, and the RL model is deployed on the base station or a single SU. In this method, the interactions among multiple SUs are ignored, which is not reasonable in practice. Naparstek and Cohen~\cite{naparstek2018deep} proposed a distributed RL approach for DSA, where an MARL algorithm is introduced with considering the interference among multiple SUs. Although RL-based techniques such as Q-learning have demonstrated good performance on DSA, a critical challenge is it fails as the user number increases. In this paper, our proposed method significantly improves the performance with massive users. 
\section{Conclusion}
\label{sec: conclusion}

In this paper, we present Hermes-- a decentralized DSA method in 5G with the consideration of massive UEs.
In Hermes, we design an improved MARL algorithm for training local models at UEs in order to achieve better resource utilization. A novel shuffle mechanism is also proposed to exchange trained models among UEs to realize better fairness.
Comprehensive simulation experiments demonstrate that Hermes performs better than the compared baseline methods, achieving 84 \% channel utilization efficiency and over 0.96 in JFI. Furthermore, it takes only 0.5 seconds for adapting to the real-time changes in the dynamic environment. 
We believe our work can facilitate the large-scale deployment of mobile devices in 5G, and the idea of shuffling can be easily extended to more decentralized applications.

\section*{Acknowledgement}
This work was supported in part by National Key R\&D Program of China (2018YFB0105000); and in part by National Natural Science Foundation of China (No. U19B2019, 61832007, 61621091); and in part by Tsinghua EE Xilinx AI Research Fund; and in part by Beijing National Research Center for Information Science and Technology (BNRist); and in part by Beijing Innovation Center for Future Chips.

\balance

\bibliographystyle{abbrv}
\bibliography{reference}

\end{document}